# Optimal Assembly of Repurposed Lithium-Ion Battery Packs under Cell Heterogeneity and Screening Uncertainty

Hassan Zahid Butt, *Student Member*, *IEEE* and Xingpeng Li, *Senior Member*, *IEEE*

***Abstract*— The growing supply of retired electric vehicle batteries presents an opportunity for second-life stationary energy storage, but assembling heterogeneous retired cells into reliable packs is challenging due to substantial variation in capacity, DC internal resistance (DCIR), and self-discharge. This paper proposes a robust optimization framework for cell-to-pack assembly of second-life batteries. A topology-screening stage first identifies minimum-cell series–parallel configurations satisfying inverter and energy requirements, reducing the dimensionality of the subsequent assignment problem. For each candidate topology, a mixed-integer linear program selects cells and assigns them along the series string, enforcing power, voltage, and energy requirements as hard constraints while minimizing a normalized, weighted sum of DCIR spread, capacity spread, and self-discharge imbalance. Additionally, measurement uncertainty in capacity and DCIR is modeled as bounded intervals to guarantee feasibility under worst-case parameter deviations. The framework is evaluated on four heterogeneous inventories for a 10 kW/10 kWh stationary backup application. The proposed method satisfies all feasibility requirements in every case, while single-metric sorting heuristics each fail on at least one inventory. Relative to the best single-metric baseline by objective value, it reduces the normalized mismatch objective by 76–87%, demonstrating that jointly optimizing cell matching with application-level feasibility requirements improves heterogeneous second-life pack assembly under screening uncertainty.**

## NOMENCLATURE

Sets and indices:

| | |
|---|---|
| $C$ | Available cells in the inventory; index $i$ |
| $G$ | Parallel groups; index $g$ |

Parameters:

| | |
|---|---|
| $Q_i$ | Capacity of cell $i$ (Ah) |
| $R_i$ | DC internal resistance of cell $i$ (mΩ) |
| $SD_i$ | Self-discharge per month of cell $i$ (%/month) |
| $\Delta Q_i$ | Capacity uncertainty band of cell $i$ (Ah) |
| $\Delta R_i$ | DCIR uncertainty of cell $i$ (mΩ) |
| $V_{cell}^{nom}$ | Nominal voltage of one cell (V) |
| $V_{cell}^{min}, V_{cell}^{max}$ | Min / max allowable cell voltage (V) |
| $Q^{min}, Q^{max}$ | Min / max cell capacity in inventory (Ah) |
| $R^{min}, R^{max}$ | Min / max DCIR value in inventory (mΩ) |
| $SD^{min}, SD^{max}$ | Min / max self-discharge in inventory (%/month) |
| $E^{req}$ | Required usable pack energy (Wh) |
| $P^{req}$ | Required battery pack power (W) |
| $T^{bk}$ | Required backup duration (h) |
| $V_{inv}^{min}, V_{inv}^{max}$ | Min / max inverter required pack voltage (V) |
| $I_{inv}^{max}$ | Maximum inverter allowable pack current (A) |
| $N_s$ | Number of series groups |
| $N_p$ | Number of parallel cells per group |
| $V^{nom}$ | Nominal pack voltage (V) |
| $V_{ld}^{min}$ | Minimum loaded pack voltage (V) |
| $M_Q, M_R$ | Big-M for capacity- and resistance-linking |
| $\gamma$ | Relative DCIR matching tolerance within group |
| $\omega_Q, \omega_R, \omega_{SD}$ | Weights for capacity, DCIR, self-discharge spread |
| $\beta_Q, \beta_R, \beta_{SD}$ | Normalization constants for objective function |

Variables:

| | |
|---|---|
| $x_{ig}$ | Binary variable, =1 if cell $i$ is assigned to group $g$ |
| $q_g, q_g^{rob}$ | Nominal / worst-case group capacity (Ah) |
| $q_{pack}, q_{pack}^{rob}$ | Nominal / worst-case pack capacity (Ah) |
| $r_g^{min}, r_g^{max}$ | Min / max nominal DCIR in group $g$ (mΩ) |
| $r_{g,rob}^{min}, r_{g,rob}^{max}$ | Min / max worst-case DCIR in group $g$ (mΩ) |
| $r_{pack}^{rob}$ | Worst-case pack resistance (mΩ) |
| $sd_g$ | Mean self-discharge of parallel group $g$ (%/month) |
| $sd_{pack}^{min}, sd_{pack}^{max}$ | Min, max group-mean self-discharge across groups (%/month) |
| $q_g^{min}, q_g^{max}$ | Min, max capacity among cells in group $g$ (Ah) |

Hassan Zahid Butt and Xingpeng Li are with the Department of Electrical and Computer Engineering, University of Houston, Houston, TX, 77204, USA. (e-mail: hbutt@uh.edu; xli82@uh.edu).

## I. INTRODUCTION

Lithium-ion batteries are central to modern electrification, and the growing supply of retired electric vehicle (EV) cells has renewed interest in second-life deployment [1]. Driven by rapid EV adoption, retired batteries available for second-life stationary use could exceed 200 GWh per year by 2030 [2], and roughly 250,000 metric tons were projected to reach end of life by 2025 [3]. Rather than recycling these cells immediately, repurposing them as second-life batteries (SLBs) extends their economic value in roles like stationary backup, peak shaving, and microgrid energy shifting [4].

Cell-to-cell variation in capacity, DC internal resistance (DCIR), and self-discharge appears even within a single production batch and widens as cells age [5] with direct pack-level consequences: resistance mismatch among parallel cells distorts current sharing and accelerates degradation [6], while self-discharge variability promotes long-term energy imbalance along the series string [7]. Since a series string is limited by its lowest-capacity group, pairing high-capacity cells with degraded ones reduces usable pack energy.

This heterogeneity challenge is further compounded by measurement uncertainty in second-life screening. Although the battery passport concept, gaining regulatory traction through initiatives such as EU Regulation 2023/1542 and the Battery Pass framework, aims to improve lifecycle data traceability [8], data completeness and consistency across repurposing stakeholders remain difficult in practice [9]. Facilities therefore rely on diagnostic screening estimates whose accuracy depends on test conditions [10], and a pack assembled against nominal values alone may fail under realistic deviations — motivating a design that remains feasible under bounded measurement error.

Even with precise cell-parameter knowledge, the assembly problem is inherently combinatorial: cell selection, series grouping, parallel grouping, and the series-parallel topology jointly determine pack energy, power output, voltage regulation, and current distribution. Because these outcomes are coupled, ranking cells by any single metric is generally insufficient to satisfy the energy, power, voltage, and current requirements simultaneously. Motivated by these challenges, this paper formulates second-life cell-to-pack assembly as a robust optimization problem combining analytic topology screening with mixed-integer linear programming (MILP) that selects cells and assigns them to parallel-group positions along the series string under bounded capacity and DCIR measurement uncertainty, guaranteeing feasibility under worst-case parameter deviations.

## II. Literature Review

Literature extensively examines SLB adoption from technical-feasibility and qualification perspectives. Studies show retired cells can satisfy stationary applications after first-life degradation [11], and NREL qualification procedures provide practical protocols for condition assessment [12], [13]. Technoeconomic analyses further show that SLB value depends strongly on degradation and that repurposing decisions should align with application-specific requirements rather than generic state-of-health thresholds [14]. While these works establish the case for SLB deployment, they do not address the downstream problem of assembling application-compliant packs from heterogeneous inventories.

A second line of research addresses cell sorting and grouping to improve pack consistency. Data-driven pipelines combining unsupervised clustering with learned estimators have been proposed to accelerate screening and reduce diagnostic testing [15]. Piombo et al. [16] introduced an application-aware regrouping approach that improves SLB pack consistency by accounting for application context during sorting – a step toward the problem considered here, though without topology flexibility or measurement uncertainty consideration. Deep learning methods have also been proposed to maintain grouping consistency under unknown prior conditions [17]. These approaches rely on similarity metrics and do not enforce pack-level design constraints such as power capability, voltage limits, or backup duration.

A smaller but growing body of work formulates SLB assembly as an optimization problem. Lee and Kum [18] proposed a cell-selection framework that tunes screening thresholds to reduce heterogeneity for a given requirement set, though cell-to-position assignment is treated as a separate downstream step. Wu et al. [19] introduced a two-step sorting and regrouping strategy that accounts for series–parallel topological structure of the battery pack, improving pack consistency beyond what single-metric sorting achieves. Dang et al. [20] investigated reconfigurable battery networks in which switching elements dynamically alter cell connectivity to compensate for heterogeneity during operation, a complementary but distinct approach that introduces switching losses and complexity, absent in fixed-topology designs.

### A. Research Gaps & Contributions

Despite these advances, three gaps remain. First, integrating topology screening, cell selection, and cell-to-position assignment against application requirements within a single formulation is underexplored. Second, application constraints like power capability, backup duration, and voltage/current limits are seldom directly integrated. Third, most assembly models treat parameters as exact, ignoring measurement uncertainty that can render a nominally feasible pack infeasible. This paper addresses these gaps with the following contributions:

i. A unified framework that screens series-parallel topologies and jointly handles cell selection and position assignment within a single formulation.
ii. A topology-screening stage that analytically eliminates infeasible configurations before optimization, followed by a MILP enforcing application requirements as hard constraints rather than post-hoc checks.
iii. A robust formulation incorporating bounded capacity and DCIR intervals to guarantee worst-case constraint feasibility while optimizing for minimal pack mismatch.

## III. Proposed Methodology

The cell-to-pack assembly problem is solved in two phases. Phase 1 is a topology-screening stage that enumerates series–parallel configurations capable of meeting the application requirements, ranked by ascending total cell count. Phase 2 solves a robust MILP for each candidate topology in that order, assigning individual cells to series-group positions, and returns the first topology that admits a feasible assembly. Because cell count is fixed once a topology is chosen, the first feasible candidate is by construction a minimum-cell feasible design; the Phase 2 objective therefore addresses only matching quality rather than pack sizing. This decomposition reduces the combinatorial burden: by fixing the topology before optimization, Phase 2 avoids searching over series-parallel configurations and instead solves a fixed-topology assignment MILP with $|C| \times N_s$ binary variables, of which $(N_s \times N_p)$ cells are selected.

### A. Phase 1: Topology Screening

Throughout this paper, a group refers to the set of $N_p$ cells connected in parallel at a single position along the series string; the pack thus comprises $N_s$ groups connected in series. For a given cell chemistry, the admissible number of series groups is bounded by the inverter voltage window:

$$N_s^{min} = \frac{V_{inv}^{min}}{V_{cell}^{min}} \quad , \quad N_s^{max} = \frac{V_{inv}^{max}}{V_{cell}^{max}} \tag{1}$$

Only integer values of $N_s$ within $[N_s^{\min}, N_s^{\max}]$ are retained. Each candidate must additionally satisfy the rated-power current limit and the minimum loaded-voltage requirement:

$$\frac{P^{req}}{N_s V_{cell}^{nom}} \leq I_{inv}^{max} \quad , \quad N_s V_{cell}^{nom} \geq V_{ld}^{min} \tag{2}$$

where $V_{ld}^{min} = max(V_{inv}^{min}, P^{req}/I_{inv}^{max})$. For each $N_s$, the parallel count $N_p$ is bounded by inventory size and a necessary energy condition:

$$N_s N_p \leq |C| \quad , \quad V^{nom} N_p Q^{max} \geq E^{req} \tag{3}$$

with $V^{nom} = N_s V_{cell}^{nom}$. Because (3) uses $Q^{max}$, the largest inventory capacity, a topology is eliminated only when the energy requirement fails even under the most optimistic selection, ensuring no realizable assembly is discarded.

### B. Phase 2: Optimization Model (OASLB)

The optimization model in this study is referred to as OASLB — Optimal Assembly of Second-Life Batteries. For a fixed topology $(N_s, N_p)$, the model assigns cells to parallel-group positions along the series string to minimize a weighted, normalized sum of three mismatch measures — DCIR spread within parallel groups, self-discharge imbalance across the series string, and capacity spread within parallel groups:

$$min\ \frac{\omega_R}{\beta_R}\sum_{g\in G}\left(r_g^{max}-r_g^{min}\right)+\frac{\omega_{SD}}{\beta_{SD}}\left(sd_{pack}^{max}-sd_{pack}^{min}\right)+\frac{\omega_Q}{\beta_Q}\sum_{g\in G}\left(q_g^{max}-q_g^{min}\right) \quad (4)$$

The constants $\beta_R = N_s(R^{max} - R^{min})$, $\beta_{SD} = SD^{max} - SD^{min}$, and $\beta_Q = N_s(Q^{max} - Q^{min})$ normalize all three terms to a common unit-free range so the weights $\omega_R$, $\omega_{SD}$, and $\omega_Q$ reflect relative priority rather than differences in physical units; equal weights are used in all case studies in this paper. Because robustness is enforced entirely through the constraints, the objective uses nominal values — optimizing matching quality under expected conditions while the constraints guarantee worst-case feasibility. Equations (5) and (6) enforce the assignment structure: each cell is assigned to at most one group, and each group contains exactly $N_p$ cells:

$$\sum_{g\in G} x_{ig} \le 1\,, \qquad \forall i \qquad (5)$$

$$\sum_{i\in C} x_{ig} = N_p\,, \qquad \forall g \qquad (6)$$

Equations (7)–(9) enforce nominal energy feasibility: (7) sums each group's capacity, (8) bounds usable pack capacity by the weakest group, and (9) requires it to meet the energy requirement at nominal voltage. Equations (10)–(12) are the robust counterparts: (10) subtracts each cell's capacity uncertainty band, (11) applies the same weakest-link bound, and (12) — the principal robustness guarantee — requires the worst-case pack capacity to meet the energy floor even when all measurements sit at the unfavorable edge of their bands:

$$q_g = \sum_{i\in C} Q_i\, x_{ig}\,, \qquad \forall g \qquad (7)$$

$$q_{pack} \le q_g\,, \qquad \forall g \qquad (8)$$

$$V^{nom} q_{pack} \ge E^{req} \qquad (9)$$

$$q_g^{rob} = q_g - \sum_{i\in C} \Delta Q_i x_{ig}\,, \qquad \forall g \qquad (10)$$

$$q_{pack}^{rob} \le q_g^{rob}\,, \qquad \forall g \qquad (11)$$

$$V^{nom} q_{pack}^{rob} \ge E^{req} \qquad (12)$$

Equations (13)–(15) track nominal DCIR spread within each group via standard Big-M linking, and (16)–(18) are the robust counterparts applied to the edges of each cell's DCIR uncertainty band. Equation (18) enforces a hard compatibility limit: the worst-case maximum DCIR in any group cannot exceed $1+\gamma$ times the worst-case minimum. The objective's DCIR spread term minimizes spread as a continuous quality measure, but it can still co-assign incompatible cells if that lowers the overall mismatch; (18) instead imposes a hard bound on the within-group resistance ratio, independent of the objective value:

$$r_g^{max} \ge R_i - M_R\left(1 - x_{ig}\right)\,, \qquad \forall i, g \qquad (13)$$

$$r_g^{min} \le R_i + M_R\left(1 - x_{ig}\right)\,, \qquad \forall i, g \qquad (14)$$

$$r_g^{max} \ge r_g^{min}\,, \qquad \forall g \qquad (15)$$

$$r_{g,rob}^{max} \ge (R_i + \Delta R_i) - M_R\left(1 - x_{ig}\right)\,, \qquad \forall i, g \qquad (16)$$

$$r_{g,rob}^{min} \le (R_i - \Delta R_i) + M_R\left(1 - x_{ig}\right)\,, \qquad \forall i, g \qquad (17)$$

$$r_{g,rob}^{max} \le (1+\gamma) r_{g,rob}^{min}\,, \qquad \forall g \qquad (18)$$

Equations (19)–(21) apply the same Big-M structure to capacity spread, where $q_g^{max}$ and $q_g^{min}$ denote the max and min individual-cell capacities in group $g$. These objective variables differ from total group capacity in (7)–(12):

$$q_g^{max} \ge Q_i - M_Q\left(1 - x_{ig}\right)\,, \qquad \forall i, g \qquad (19)$$

$$q_g^{min} \le Q_i + M_Q\left(1 - x_{ig}\right)\,, \qquad \forall i, g \qquad (20)$$

$$q_g^{max} \ge q_g^{min}\,, \qquad \forall g \qquad (21)$$

Equations (22)–(24) address self-discharge balance. Cells in a parallel group share a terminal voltage, so the group's self-discharge is represented by the mean rate of its assigned cells, defined by (22). Since group capacities are closely matched in the studied inventories, the group-mean self-discharge rate is used as a linear approximation of the capacity-weighted self-discharge rate. (23)–(24) identify the max and min group-mean self-discharge across the string, whose difference the objective minimizes to limit long-term SOC divergence:

$$N_p sd_g = \sum_{i\in C} SD_i x_{ig}\,, \qquad \forall g \qquad (22)$$

$$sd_{pack}^{max} \ge sd_g\,, \qquad \forall g \qquad (23)$$

$$sd_{pack}^{min} \le sd_g\,, \qquad \forall g \qquad (24)$$

Equations (25) and (26) enforce the loaded-voltage requirement under worst-case conditions. To preserve MILP linearity, each parallel group's resistance is approximated by a conservative linear expression — the mean worst-case cell DCIR over $N_p$ — summed over the $N_s$ groups to give the worst-case pack resistance; the $N_p^2$ denominator in (25) follows from this parallel-series combination. The approximation coincides with the exact parallel resistance when within-group DCIR values are equal and remains tight under the 10% matching tolerance imposed by (18). Equation (26) requires the pack terminal voltage under rated load — accounting for resistive drop, the factor 1000 converting mΩ to Ω — to stay above the loaded-voltage floor:

$$r_{pack}^{rob} = \sum_{g\in G} \frac{\sum_{i\in C}(R_i + \Delta R_i) x_{ig}}{N_p^2} \qquad (25)$$

$$V^{nom} - \frac{I_{inv}^{max} r_{pack}^{rob}}{1000} \ge V_{ld}^{min} \qquad (26)$$

Equation (27) breaks symmetry by ordering groups according to non-decreasing nominal minimum DCIR, eliminating $N_s!$ redundant label permutations. Valid-inequality cuts on cell pairs violating (18) tighten the LP relaxation to reduce solver runtime.

$$r_g^{min} \le r_{g+1}^{min}\,, \qquad g = 1, \ldots, N_s - 1 \qquad (27)$$

Table I summarizes the three assembly methods compared. Deterministic and robust OASLB solve the same fixed-topology MILP and differ only in the uncertainty setting: the deterministic case sets $\Delta Q_i = \Delta R_i = 0$ and is evaluated at nominal screened values, whereas the robust case uses the bounded uncertainty intervals and is evaluated under worst-case capacity and DCIR deviations. The three sorting baselines are non-optimization heuristics that enforce no pack-level constraints: cells are ranked by a single attribute and adjacent cells are grouped, after which each assembly is post-evaluated under the same worst-case feasibility checks. A baseline is therefore labeled infeasible when its assignment violates the energy-floor, DCIR-tolerance, or loaded-voltage check, even though no constraint prevented that assignment.

TABLE I: ASSEMBLY METHODS COMPARED IN THIS STUDY

| Method | Uncertainty | In-model feasibility |
|---|---|---|
| **OASLB (deterministic)** | $\Delta Q_i = \Delta R_i = 0$ | Enforced (nominally) |
| **OASLB (robust)** | Bounded intervals | Enforced (worst-case) |
| **Single-metric sorting** | Not modeled | Post-checked only |

## IV. CASE STUDIES

The proposed framework is evaluated on four controlled lithium-iron-phosphate (LFP) cell inventories of 100 cells each, with nominal voltage 3.2 V and capacity centered around 100 Ah. Datasets are generated using truncated Gaussian distributions with fixed seeds for reproducibility. The four inventories are designed to stress distinct aspects of the problem, as summarized in Table II, where mean and standard deviation (std) denote the target distribution parameters used to generate each 100-cell inventory. D1–D3 isolate and combine distinct heterogeneity sources: D1 isolates DCIR-matching with tight capacity spread, D2 stresses weakest-link energy with wide capacity and tight DCIR dispersion, and D3 couples both while additionally widening self-discharge dispersion. D4 introduces wider measurement uncertainty bands to stress the robust formulation against poorly calibrated screening. The datasets differ mainly in capacity dispersion, DCIR dispersion, uncertainty level, and, for D3, self-discharge dispersion.

TABLE II: INVENTORY PARAMETERS (100 LFP CELLS EACH)

| | D1 | D2 | D3 | D4 |
|---|---|---|---|---|
| **Cap. mean/std (Ah)** | 100/0.6 | 100/4.0 | 100/4.0 | 100/2.0 |
| **DCIR mean/std (mΩ)** | 0.55/0.05 | 0.55/0.02 | 0.55/0.05 | 0.55/0.035 |
| **SD mean/std (%/mo.)** | 2.0/0.4 | 2.0/0.4 | 2.0/0.6 | 2.0/0.4 |
| **Cap. uncertainty (±%)** | 0.5 | 0.5 | 0.5 | 2.0 |
| **DCIR uncertainty (±%)** | 1.0 | 1.0 | 1.0 | 3.0 |

The target application is a 10 kW/10 kWh stationary backup system unless stated otherwise, so the usable-energy requirement is $E^{\text{req}} = P^{\text{req}} \cdot T^{\text{bk}}$ with $T^{\text{bk}} = 1$ h. The pack voltage window is 40–60 V, the maximum pack current is 200 A, and the loaded-voltage floor is 50 V — set by the rated-power current limit. The DCIR matching tolerance $\gamma$ is fixed at 10% throughout. Three single-metric sorting heuristics serve as baselines: capacity-sorted (descending), DCIR-sorted (ascending), and self-discharge-sorted (ascending). For each method, cells are sorted globally on the chosen metric and adjacent cells in the sorted order are assigned to the same parallel group, which minimizes within-group spread on that metric. The self-discharge baseline tests whether sorting only by degradation-related imbalance can also preserve energy and DCIR feasibility. All methods are evaluated under the same worst-case feasibility checks: energy floor, loaded voltage, and DCIR tolerance, and a method is considered feasible only if it satisfies all three simultaneously. The case studies are organized into three test scenarios: TS1 examines topology selection and sizing, verifying that the screening stage identifies the minimum-cell topology for a given application; TS2 compares deterministic and robust assemblies across all four inventories; and TS3 benchmarks the robust OASLB solution against the three sorting baselines.

### A. *TS1: Topology Selection and Sizing*

The first study verifies topology screening selects the minimum-cell topology satisfying application requirements. Dataset D1 is used because its tight capacity distribution isolates the effect of power and energy requirements on topology selection. For the 5 kW/5 kWh case, the screening stage selects a 16S1P topology, where 16S1P denotes $N_s = 16$ series groups and $N_p = 1$ cell per group, using 16 cells and delivering 5.07 kWh at a loaded voltage of 50.09 V with a returned objective value of 0.039. For the 10 kW/10 kWh case, the 16S1P topology is insufficient to meet the coupled energy and current requirements, and the framework advances to 16S2P, using 32 cells and delivering 10.14 kWh at a loaded voltage of 50.35 V. In both cases the selected topology satisfies the application requirements with approximately 1.4% energy excess — confirming that the screening stage selects right-sized topologies and that parallel count is determined by application power and energy requirements, not chosen arbitrarily.

### B. *TS2: Effect of Robustness under Screening Uncertainty*

The second study compares deterministic and robust assemblies for the 10 kW/10 kWh application on all four inventories using the 16S2P topology. Fig. 1 summarizes the results across three metrics: mismatch objective, worst-case pack energy, and worst-case DCIR spread.

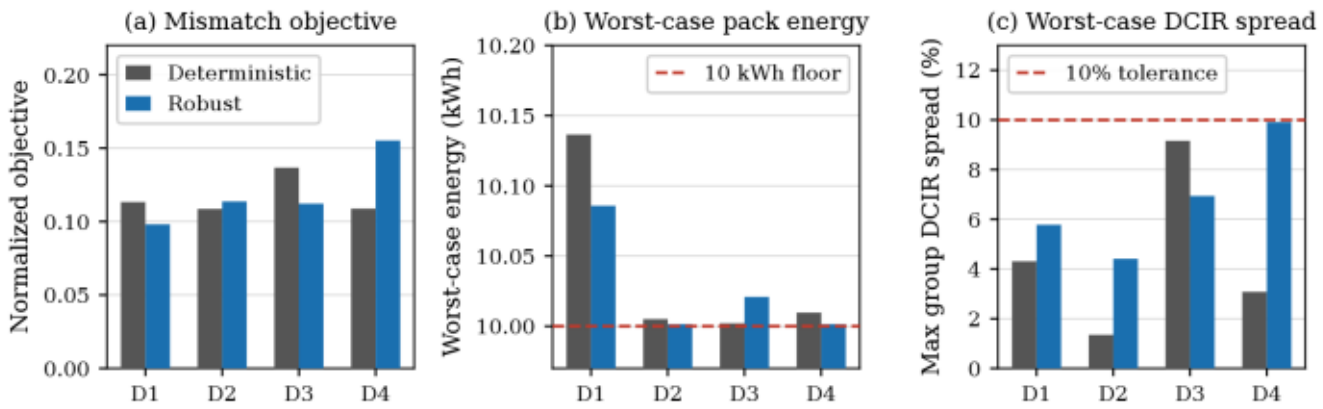

Fig. 1. Deterministic and robust assemblies across four inventories: (a) mismatch objective, (b) pack energy, and (c) maximum group DCIR spread.

As shown in Fig. 1(b), the robust assembly satisfies the 10-kWh worst-case energy floor in every inventory, with margins ranging from 1.5 Wh in D2/D4 to 85.8 Wh in D1. The deterministic assembly also meets the nominal floor, but because it applies no uncertainty margin, its reported energy equals its nominal value and cannot guarantee feasibility under measurement error. Fig. 1(c) shows that the worst-case DCIR spread remains below the 10% tolerance in all robust cases, with the tightest margin in D4 (9.91%), where wider uncertainty bands restrict feasible cell combinations. The deterministic D3 case approaches the tolerance limit at 9.16% nominally, with no robust margin. The robust objective remains close to the deterministic one across inventories—comparable in D1 and D3, and higher in D4, where it rises from 0.109 to 0.155 as wider uncertainty bands tighten the effective DCIR and capacity constraints, leaving fewer well-matched combinations. Thus, robustness provides explicit worst-case feasibility guarantees at only a modest, inventory-dependent cost in nominal matching quality.

### C. *TS3: Comparison with Single-Metric Sorting Heuristics*

The third study compares the robust OASLB solution against the three sorting baselines for the 10 kW/10 kWh, 16S2P case. Fig. 2 presents the normalized mismatch objective and worst-case pack energy across all four inventories. A method is considered infeasible if it violates any of the three requirements — energy floor, DCIR tolerance, or loaded-voltage limit — under worst-case parameter evaluation; infeasible bars are indicated by hatching in Fig. 2.

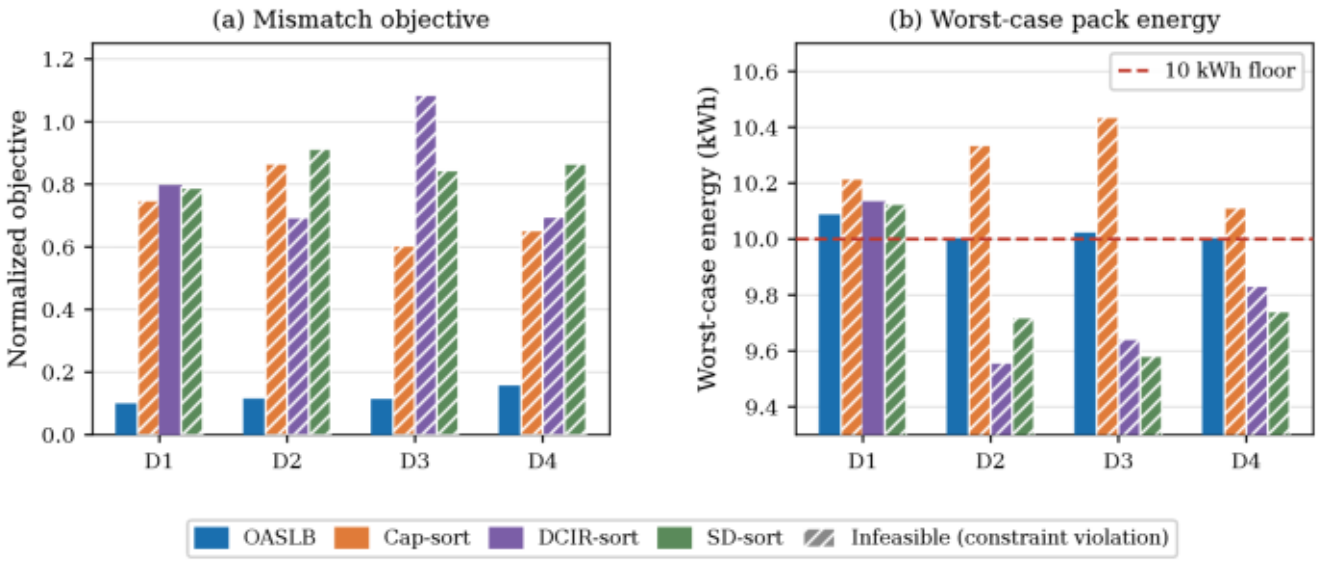

**Fig. 2.** Comparison of robust OASLB with single-metric sorting baselines.

OASLB is the only method that satisfies all three requirements across all four inventories. Relative to the lowest-objective sorting baseline on each inventory — regardless of whether that baseline is itself feasible — OASLB reduces the normalized mismatch objective by 86.9% (D1), 83.6% (D2), 81.4% (D3), and 76.2% (D4), as shown in Fig. 2(a). The pattern of baseline failures reveals the structural limitation of single-metric sorting. Capacity sorting selects high-energy cells and satisfies the energy floor in all four inventories, but assigns cells without regard to DCIR compatibility, resulting in worst-case group spreads reaching 28.3% in D1 and 32.9% in D3 — far exceeding the 10% tolerance — and is therefore infeasible in all four cases. DCIR sorting avoids resistance mismatches but ignores weakest-link capacity effects, causing the worst-case pack energy to fall below 10 kWh in D2 (9.56 kWh), D3 (9.64 kWh), and D4 (9.83 kWh), as visible in Fig. 2(b). Self-discharge sorting performs poorly on both axes simultaneously, incurring DCIR violations and energy-floor failures in three of the four inventories. These failures confirm that no single cell attribute captures the coupled effects of energy sufficiency, current-sharing compatibility, voltage sag, and series-string imbalance, highlighting the value of enforcing these coupled requirements jointly through the assembly optimization.

## V. Conclusion

This paper presented a robust mixed-integer optimization framework for cell-to-pack assembly of second-life batteries from heterogeneous inventories. The proposed framework pairs a minimum-cell topology-screening stage with a MILP that jointly selects cells and assigns them along the series string, enforcing power, backup-duration, voltage, and current requirements as hard constraints with capacity and DCIR uncertainty as bounded intervals to guarantee worst-case feasibility. The framework was evaluated on four synthetic inventories spanning DCIR-dominant, capacity-dominant, doubly heterogeneous, and measurement-uncertainty-stress conditions for a 10 kW/10 kWh stationary backup application. In all four inventories, the robust formulation simultaneously satisfied the worst-case energy floor, DCIR tolerance, and loaded-voltage requirements, whereas each single-metric sorting heuristic — capacity-, DCIR-, and self-discharge-sorted — violated at least one of these requirements. Relative to the best-performing single-metric baseline by objective value on each inventory, the proposed method reduced the normalized mismatch objective by 76–87%, demonstrating that cell heterogeneity and measurement uncertainty can be effectively addressed as integral components of the assembly formulation. Future work will focus on validating the proposed framework on field-measured retired-cell inventories to assess its practical performance under real-world parameter distributions and screening conditions.

## References

[1] E. Martinez-Laserna, I. Gandiaga, E. Sarasketa-Zabala, J. Badeda, D.-I. Stroe, M. Swierczynski, and A. Goikoetxea, "Battery second life: Hype, hope or reality? A critical review of the state of the art," *Renewable and Sustainable Energy Reviews*, vol. 93, pp. 701–718, 2018.

[2] H. Engel, P. Hertzke, and G. Siccardo, "Second-life EV batteries: The newest value pool in energy storage," McKinsey & Company, Apr. 2019. [Online]. Available: www.mckinsey.com/industries/automotive-and-assembly/our-insights/second-life-ev-batteries-the-newest-value-pool-in-energy-storage

[3] H. Song, H. Chen, Y. Wang, and X. E. Sun, "An overview about second-life battery utilization for energy storage: Key challenges and solutions," Energies, vol. 17, no. 23, 2024.

[4] H. Z. Butt and X. Li, "Planning future microgrids with second-life batteries: A degradation-aware iterative optimization framework," arXiv:2503.10900, 2025.

[5] K. Rumpf, M. Naumann, and A. Jossen, "Experimental investigation of parametric cell-to-cell variation and correlation based on 1100 commercial lithium-ion cells," J. Energy Storage, vol. 14, pp. 224–243, 2017.

[6] R. Gogoana, M. B. Pinson, M. Z. Bazant, and S. E. Sarma, "Internal resistance matching for parallel-connected lithium-ion cells and impacts on battery pack cycle life," J. Power Sources, vol. 252, pp. 8–13, 2014.

[7] I. Zilberman, S. Ludwig, and A. Jossen, "Cell-to-cell variation of calendar aging and reversible self-discharge in 18650 nickel-rich, silicon–graphite lithium-ion cells," J. Energy Storage, vol. 26, p. 100900, 2019.

[8] European Parliament, Regulation (EU) 2023/1542 of the European Parliament and of the Council Concerning Batteries and Waste Batteries, 2023. [Online]. Available: https://eur-lex.europa.eu/eli/reg/2023/1542/oj

[9] M. Terkes, A. Demirci, E. Gokalp, and U. Cali, "Battery passport for second-life batteries: Potential applications and challenges," IEEE Access, vol. 12, pp. 128424–128467, 2024.

[10] G. dos Reis, C. Strange, M. Yadav, and S. Li, "Lithium-ion battery data and where to find it," Energy and AI, vol. 5, p. 100081, 2021.

[11] E. Martinez-Laserna et al., "Technical viability of battery second life: A study from the ageing perspective," IEEE Trans. Ind. Appl., vol. 54, no. 3, pp. 2703–2713, May–Jun. 2018.

[12] T. Montes et al., "Procedure for assessing the suitability of battery second life applications after EV first life," Batteries, vol. 8, no. 9, p. 122, 2022.

[13] J. Neubauer, K. Smith, E. Wood, and A. Pesaran, "Identifying and overcoming critical barriers to widespread second use of PEV batteries," National Renewable Energy Laboratory, Golden, CO, Tech. Rep. NREL/TP-5400-63332, 2015.

[14] I. Mathews et al., "Technoeconomic model of second-life batteries for utility-scale solar considering calendar and cycle aging," Appl. Energy, vol. 269, p. 115127, 2020.

[15] A. Ran et al., "Fast clustering of retired lithium-ion batteries for secondary life with a two-step learning method," ACS Energy Lett., vol. 7, no. 11, pp. 3817–3825, 2022.

[16] G. Piombo, M. Faraji Niri, and J. Marco, "A novel application-aware retired lithium-ion batteries regrouping approach to enable improved second life," J. Cleaner Prod., vol. 489, p. 144526, 2025.

[17] S. Wang, F. Gao, and H. Tian, "Deep sorting of reused batteries for enabling long-term consistency grouping with unknown prior conditions," Cell Rep. Phys. Sci., vol. 6, no. 7, p. 102657, 2025.

[18] K. Lee and D. Kum, "Development of cell selection framework for second-life cells with homogeneous properties," Int. J. Electr. Power Energy Syst., vol. 105, pp. 429–439, 2019.

[19] Q. Wu, W. Ye, R. Liang, R. Pan, and T. Ma, "Two-step sorting and regrouping method of retired lithium-ion battery cells based on branches' topological structures," Energy Fuels, vol. 38, pp. 21122–21133, 2024.

[20] J. Dang, D. Xiao, X. Zhang, R. Jia, and Y. Jiao, "Optimal configuration of retired battery reconfigurable network considering switching losses," J. Energy Storage, vol. 101, p. 113735, 2024.